\documentclass[conference,compsoc]{IEEEtran}
\pdfoutput=1
\usepackage{url}
\usepackage[utf8]{inputenc}
\usepackage{wasysym}
\usepackage{xspace}
\usepackage{graphicx}

\newcommand{\system}{MOS\xspace}
\newcommand{\kernel}{BioKernel\xspace}

\begin{document}
\pagenumbering{arabic}
\pagestyle{plain}

\title{Security, extensibility, and redundancy in the Metabolic Operating System}
\author{Samuel T. King \\ University of California, Davis}

\maketitle

\thispagestyle{plain}

\begin{abstract}
People living with Type 1 Diabetes (T1D) lose the ability to produce
insulin naturally. To compensate, they inject synthetic insulin. One
common way to inject insulin is through automated insulin delivery
systems, which use sensors to monitor their metabolic state and an
insulin pump device to adjust insulin to adapt.

In this paper, we present the Metabolic Operating System, a new
automated insulin delivery system that we designed from the ground up
using security first principles. From an architecture perspective, we
apply separation principles to simplify the core system and isolate
non-critical functionality from the core closed-loop algorithm. From
an algorithmic perspective, we evaluate trends in insulin technology
and formulate a simple, but effective, algorithm given the
state-of-the-art. From a safety perspective, we build in multiple
layers of redundancy to ensure that the person using our system
remains safe.

Fundamentally, this paper is a paper on real-world experiences
building and running an automated insulin delivery system. We report
on the design iterations we make based on experiences working with one
individual using our system. Our evaluation shows that an automated
insulin delivery system built from the ground up using security first
principles can still help manage T1D effectively.

Our source code is open source and available on GitHub (link omitted).

\end{abstract}

\section{Introduction}
\textbf{Biohacker} /'bīō,haker/ \textit{Noun}
\begin{enumerate}
  \item A person who manipulates their metabolic state using sensors,
    injected hormones, nutrients, physical activity, computer systems,
    and artificial intelligence.
  \item An enthusiastic and curious person who learns about their own
    biology and metabolism through experimentation on them self.
  \item A person who uses computers to gain access to someone's
    metabolic state.
\end{enumerate}

There are 8.4 million people living with Type 1 Diabetes (T1D)
worldwide \cite{t1d_numbers} and they are all biohackers.

T1D is a metabolic disorder where people's immune systems attack their
pancreas and kill the cells that produce insulin. Insulin is a hormone
that transports glucose (sugar) from the blood stream to muscles, the
liver, the brain, and other places where the body uses it for
energy. When people eat, their digestive system converts food into
glucose, which makes its way into their blood stream. Since people
with T1D are unable to produce their own insulin, they inject
synthetic insulin, and they play the role of the pancreas in their
metabolic system.

Managing T1D is hard. First, each time a person living with T1D eats,
they need to make their own dynamic dosing decisions
\cite{scheiner2020think}. This process requires considering their
current metabolic state, food, insulin, exercise, stress, caffeine,
and so on to come up with an accurate prediction for how their blood
glucose will respond to the meal and injected insulin.  Then, they
need to match the timing of the insulin injection with the timing of
when food digests and causes glucose to enter the blood
stream. Finally, after the process of injecting insulin and eating is
done, they need to track their blood glucose level for hours afterward
to make sure that they got it right and adjust if they got something
wrong. And this process repeats every time they eat -- meals, snacks,
desert, anything -- causing a substantial cognitive load to manage
T1D.

Second, insulin is a dangerous hormone and can kill people. Taking
too much insulin can kill people in a matter of hours
\cite{cryer2012severe} and taking too little insulin can kill people
in a matter of days \cite{benoit2018trends}. People with T1D need to
walk a delicate balance with insulin dosing and food to avoid serious
consequences.

Third, after people living with T1D learn how to \textit{not} die,
even less severely imbalanced glucose levels can still cause long-term
health complications. These complications include heart disease,
stroke, kidney failure, blindness, nerve damage, amputations, and
impotence (in men), just to name a few \cite{diabetes1993effect,
  genuth2006insights}. The best way to avoid these long-term health
complications is to manage tight control of glucose levels.

Technology and biohacking help to manage T1D. Changes in diet can help
provide more predictable glucose responses after eating
\cite{lennerz2018management}. Continuous Glucose Monitors (CGMs) track
glucose levels using implanted sensors to facilitate real time
treatment adaptations \cite{ponder2019sugar}. And automated insulin
delivery software can connect these CGM readings with an insulin pump
for automatic insulin dosing \cite{bequette2005critical,
  cobelli2011artificial, beta_bionics}.

Biohacking presents both a challenge and an opportunity for systems
software. The challenge is that stakes are high, people's lives and
long-term health depend on the software they use for biohacking when
managing T1D, so simplicity, security, and correctness are all
critical. If we can solve this challenge, the opportunity is to create
extensibility mechanisms so that people can add to biohacking software
safely and create an ecosystem of apps that accelerate innovation for
managing T1D.

We introduce a new system, the Metabolic Operating System (\system),
that we built for biohacking and managing T1D. Our innovation is in
the architecture we use to implement T1D management features. We
use separation principles from the OS and microkernel areas
\cite{accetta1986mach, hartig1997performance} applied to the
application layer for strong isolation and simplicity of our software
components, similar to secure web browsers \cite{grier2008secure,
  wang2009multi, tang2010trust, reis2019site}. In our design, we
define biohacking abstractions, we decompose the system into isolated
modules, and expose narrow and well-defined interfaces. These
interfaces help provide the anchor for our security policies and form
the foundation for our extensibility mechanisms.

At the heart of our design is our \emph{\kernel}, which manages the
CGM and insulin pump hardware that we use in \system. The \kernel also
runs the novel closed-loop algorithm we design for automated insulin
delivery. Finally, the \kernel produces the event logs that other apps
use to infer the state of the system, which is the core of our
extensibility abstraction.

To demonstrate extensibility, we build four apps on top of
\system. Our apps handle the full life cycle of T1D management: food
entry, therapy settings ML and analysis, a replay app, and a Metabolic
Watchdog that monitors the person's metabolic state to predict high or
low glucose levels. These apps all run in isolation, consume log data,
and interact with the \kernel using a trusted UI.

Using the Metabolic Watchdog app on one individual, we show a
clinically significant improvement to their core metabolic health
metrics. We report on one year's worth of data, including four months of
treatment using \system and the Metabolic Watchdog. Our results show
an improvement in the individual's Glucose Management Index
\cite{bergenstal2018glucose}, a core measure of metabolic health for
people living with T1D. Their Glucose Management Index went from 6.8\%
at its peak to 5.8\% using \system, which is 2.6x greater than the
average improvement experienced by adults switching to fully automated
insulin delivery systems \cite{brown2021multicenter}.

Using the lessons learned from the Metabolic Watchdog, we design and
implement an automated insulin delivery system for injecting insulin
automatically using software. We report on one week's worth of data
for one individual and our results show that we maintain the tight
control we achieved with the Metabolic Watchdog while reducing the
cognitive load of managing T1D.

To the best of our knowledge, our contributions are:
\begin{itemize}
  \item \system is the first system to improve security for biohacking
    software by applying separation principles and redundancy for a
    practical system.
  \item We take a clean-slate approach to automated insulin delivery
    systems and build a new system to show how we can keep our
    implementation simple while still providing the ability to manage
    T1D.
  \item We implement a novel T1D treatment app on \system, the
    Metabolic Watchdog, and a closed-loop insulin delivery system and
    demonstrate their effectiveness on one individual.
\end{itemize}

\section{Broader trends}
Two broader trends suggest that now is the time for the research
community to invest in systems for biohacking. First, metabolic
disorders are an epidemic worldwide, with more than 133 million people
who have Prediabetes or Type 2 Diabetes in the US alone
\cite{cdc_type2}, which brings with it severe health consequences
\cite{defronzo2015type}. Although our focus is on T1D, we believe that
\system presents a first step towards software for managing metabolic
disorders in general using technology and biohacking securely.

Second, CGM technology for monitoring glucose levels are invasive
today, but research has shown that non-invasive sensors have the
potential for use in practice. Sensors for measuring glucose levels
using tears \cite{baca2007tear}, sweat \cite{moyer2012correlation},
and optical sensors \cite{steiner2011optical} all have the potential
to make available real time glucose measurements without needing to
place sensors beneath the skin, as is required with today's CGM
technology. With these non-invasive sensors we anticipate broader use,
and with broader use we will need software to help people manage their
metabolic health securely, which is the focus of this paper.

\section{Problem statement, threat model, and assumptions}
We address the problem of designing and implementing secure iOS apps
for controlling hardware that people use to manage T1D. This hardware
includes CGM sensors for reading real time glucose levels and an
insulin pump for injecting insulin. Our main security goals are
integrity and always having safe fallback states.

In our threat model, we consider attacks originating from other apps
and from network attackers. We assume that the underlying operating
system is secure and upholds its stated isolation abstractions. We
also assume that the CGM and insulin pump devices we use are correct
and that the Bluetooth pairing process establishes a secure wireless
channel between these devices and the phone. Availability attacks at
the Bluetooth or iOS level are out of scope for this paper, but we do
configure the insulin pump to have safe defaults in case availability
attacks do occur.

\section{\system design and implementation}
This paper describes our design for \system, a system for trustworthy
automated insulin delivery; we have three primary goals.  First, we
want the software that interacts with the insulin pump and CGM to be
simple. With simplicity comes the ability to reason about its
correctness and reduces the likelihood of bugs or
vulnerabilities. Second, despite this simplicity we want to support
rich functionality overall, in line with what existing automated
insulin delivery systems support today. Third, we should have the
ability to monitor the health of both our software and the human using
it.

This section describes our design and implementation for \system that
strives to achieve these goals. We lay out the principles that guide
our design, describe the overall architecture, and then discuss each
of the main components. Section \ref{sec:algorithms} describes the
algorithm design and implementation for managing T1D, this section
focuses on the systems and software architecture.

\subsection{Design principles}
In our overall design we decompose the task of automated insulin
delivery into several isolated components and provide abstractions for
these components to communicate. Together these components make up the
overall \system system. Three principles guide our design:
\begin{enumerate}
  \item \textit{Push complexity to non-critical components.} In our
    system, the software that interacts with the CGM and insulin pump
    hardware is the most critical. Thus, we push complexity away from
    this core software as much as possible to less critical
    subsystems.
  \item \textit{Ensure that the pump is always in a safe state.} This
    principle guides both how we interact with the insulin pump
    automatically and how we handle the case where the pump becomes
    disconnected from the phone.
  \item \textit{Use the longer timescales of biological systems to
    simplify software.} Biological systems operate on longer
    timescales (O(hours)) than typical computer systems. We use these
    longer timescales to look for opportunities to simplify our
    software further when we can do so safely.
\end{enumerate}

\subsection{Overall architecture}

\begin{figure}[t]
\centering
\includegraphics[width=\columnwidth]{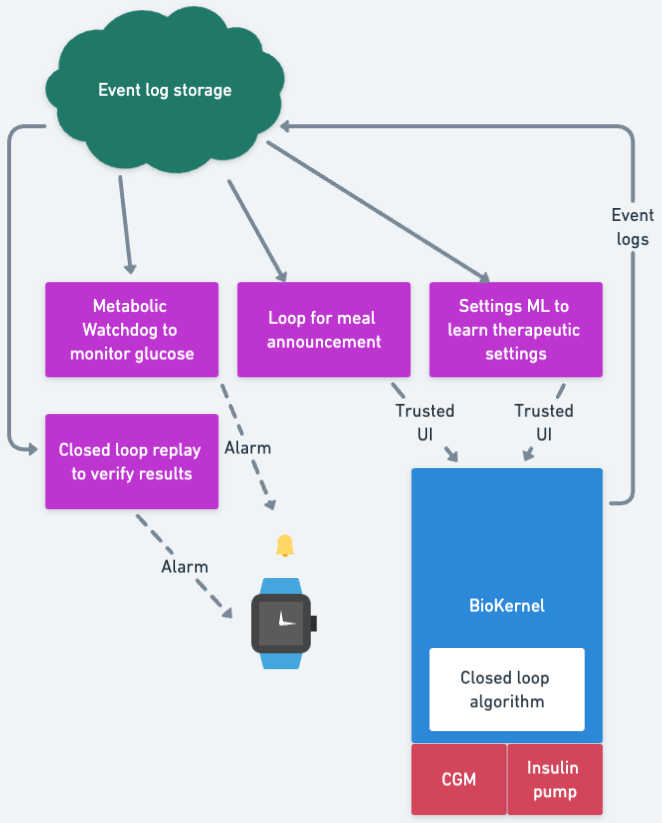}
\caption{Overall architecture for \system.}
\label{fig:architecture}
\hrulefill
\end{figure}

Figure \ref{fig:architecture} shows our overall system
architecture. At the core of our overall architecture is an iOS app,
called the \kernel. The \kernel is the component that interacts with
the CGM and insulin pump hardware, runs the closed-loop dosing
algorithm, and produces event logs that other components use to learn
the state of the system. Although the \kernel can run as a stand-alone
app that implements automated insulin delivery, we provide only a bare
minimum set of features in this core app to keep it simple.

To support rich functionality, the \kernel stores its event logs
in a cloud-based service, which is then consumed by other apps, and
exposes trusted UI views. Among the other apps that use these event
logs are a Metabolic Watchdog for monitoring the individual's
metabolic state, an app for replaying the closed-loop algorithm
execution to verify its results, an app (Loop) for meal announcements
and manual insulin dosing, and an app for running machine learning on
the data to update therapeutic settings. Apps dose insulin and update
settings through trusted UI components running within the \kernel app.

\begin{figure}[t]
\centering
\includegraphics[width=\columnwidth]{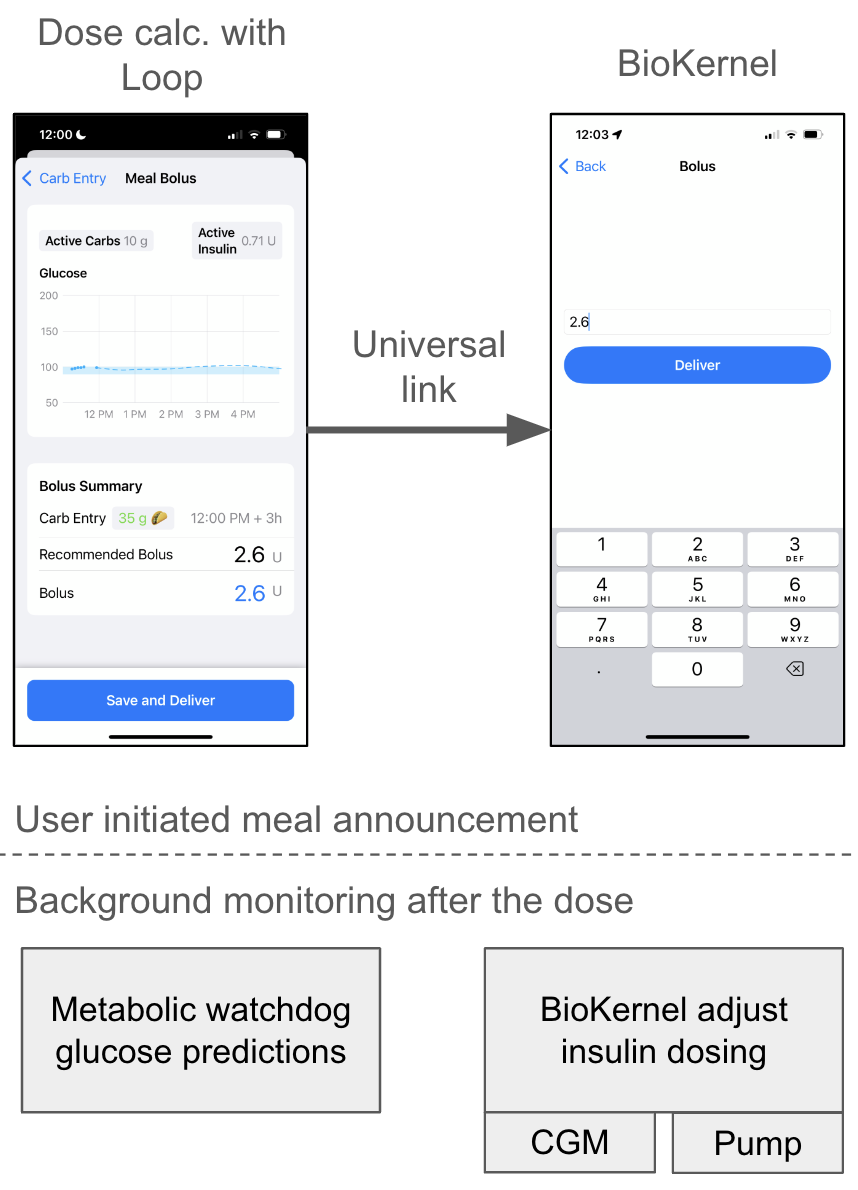}
\caption{Dosing (bolusing) for a meal example.}
\label{fig:dosing}
\hrulefill
\end{figure}

Figure \ref{fig:dosing} shows how some of these components interact
while the individual using \system takes an insulin dose before eating
a meal (called a \textit{bolus}). First, the individual opens the Loop
app, which gets the latest glucose and insulin state from event
logs. The individual enters the number of carbohydrates they will eat,
and based on this data Loop recommends an insulin dose. Once the
individual chooses an insulin dose, they press the ``Save and deliver''
button. Second, the button press uses an iOS Universal Link to display
the \kernel's insulin dosing view, which shows the individual how much
insulin it is going to deliver and gives them an opportunity to make
changes. After they press the ``Deliver'' button, the \kernel
runs a FaceID check and delivers the insulin by programming the
insulin pump.

After the initial dose, the \kernel and the Metabolic Watchdog monitor
and adjust as needed. The \kernel reads in the latest glucose readings
from the CGM in the background and adjusts the insulin dosing as
needed to maintain glycemic balance. Glycemic balance means that the
individual's glucose stays between 70 mg/dl and 140 mg/dl -- not too
high and not too low. The Metabolic Watchdog monitors glucose levels
using event logs to predict glycemic imbalance, alerting the
individual if it detects anything. Section \ref{sec:algorithms}
details the algorithms and mechanisms we use for adjusting insulin
dosing and predicting glycemic imbalance.

One implementation decision worth noting is storing our event logs in
a cloud-based service. In our original design, everything ran on
device for both improved privacy and availability. However, iOS is a
general-purpose OS and its current abstractions mismatched what we
needed in terms of background execution (see Section
\ref{subsec:ios_experience} for more details). Thus, we view the
cloud-based event logs as an artifact of our current implementation as
opposed to a fundamental aspect of our overall architecture. With the
right OS-level interfaces, we believe that everything can run on
device.

\subsection{The \kernel}

At the center of the \system is our \kernel. The \kernel manages the
CGM and insulin pump hardware, runs the closed-loop insulin dosing
algorithms, and saves event logs to the event log service to enable
other apps to recreate its state and extend its functionality.

This section provides an overview of the \kernel UI and subsystems
and discusses how we keep our implementation simple.

\subsubsection{The \kernel UI and subsystems}

\begin{figure}[t]
\centering
\includegraphics[width=0.6\columnwidth]{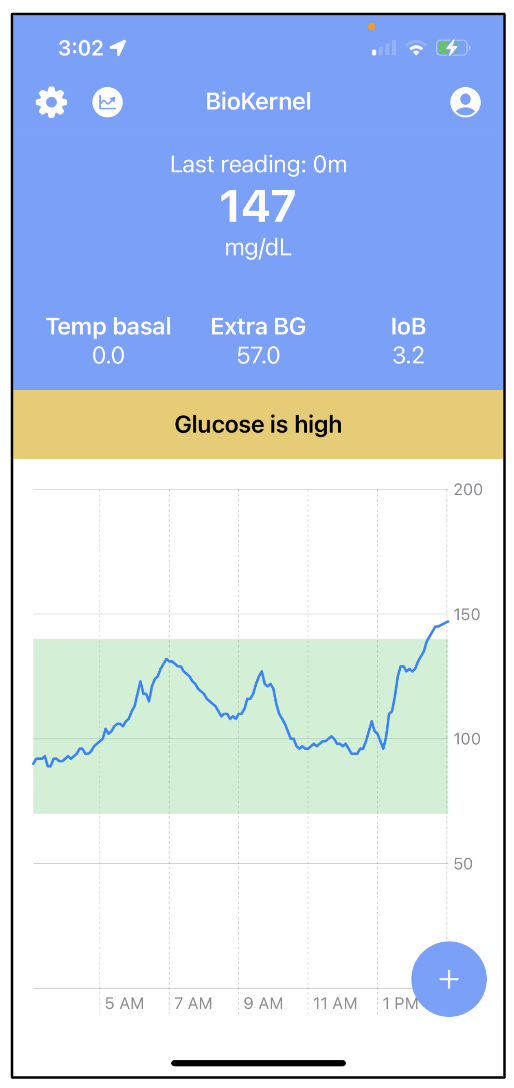}
\caption{The main user interface for the \kernel.}
\label{fig:biokernel_ui}
\hrulefill
\end{figure}

Although our design and implementation support extensibility, the
\kernel can serve as a stand-alone automated insulin delivery
system. Figure \ref{fig:biokernel_ui} shows the main UI for the
\kernel, which provides the most recent glucose reading from the CGM,
diagnostic information, an alarm interface, and a 12-hour chart of the
individual's glucose readings. From this main UI, the individual can
configure their therapeutic settings, change their pump or CGM, view
summary statistics of recent glucose readings, or deliver an insulin
dose.

\begin{figure}[t]
\centering
\includegraphics[width=\columnwidth]{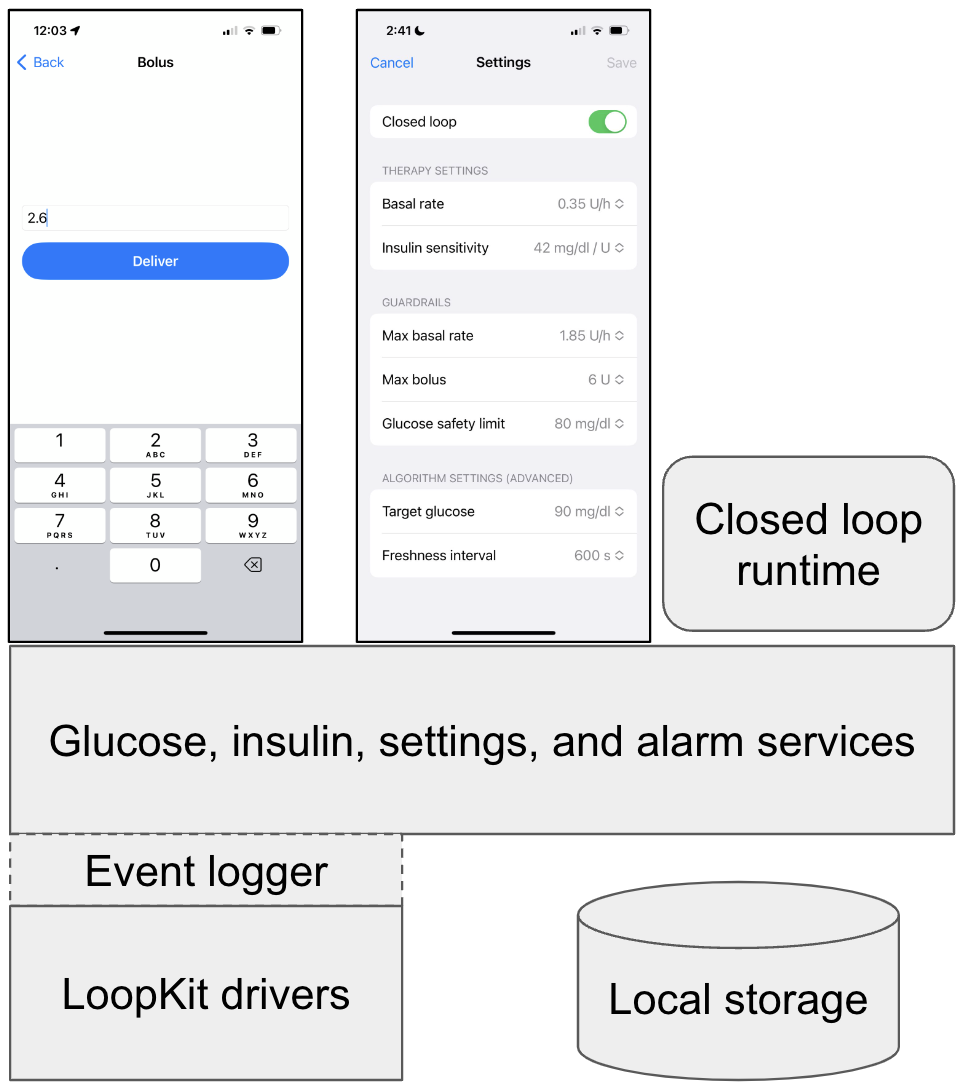}
\caption{The primary \kernel internal subsystems.}
\label{fig:biokernel}
\hrulefill
\end{figure}

Behind the scenes, the \kernel consists of five main subsystems
(Figure \ref{fig:biokernel}). First, the \kernel uses LoopKit
\cite{loopkit}, an open-source library, for device drivers to interact
with the underlying CGM and insulin pump hardware. Second, the \kernel
implements glucose, insulin, therapeutic settings, and alarm services
that ingest events from the LoopKit drivers and provide abstractions
to the rest of the \kernel. Third, the closed loop run time queries the
abstraction services and runs through the closed-loop calculation to
update insulin delivery. Fourth, the \kernel provides trusted UI
components that other apps can invoke using Universal Links to update
therapeutic settings or to dose insulin. Fifth, the \kernel event
logger sits in between the LoopKit drivers and \kernel abstraction
services to log these events.

\subsubsection{Keeping the \kernel simple}

The most important aspect of our design and implementation is keeping
the \kernel's implementation simple. With simplicity it is easier to
reason about correctness and we reduce the likelihood of bugs and
vulnerabilities. In this section, we highlight the main mechanisms we
use to keep the \kernel's implementation to a minimum. Section
\ref{subsec:biokernel_complexity} quantifies our efforts to simplify
the \kernel.

The most impactful mechanism we use for simplicity is logging events and
providing other apps with interfaces to interact with the
\kernel. Through this extensibility mechanism, we can support the
features that one would expect in an automated insulin delivery system
while keeping the \kernel itself relatively simple.

The next impactful mechanism for simplicity is through rethinking core
closed-loop algorithms with the recent advent in ultra-fast acting
insulin. Section \ref{subsec:closed_loop_algorithm_motivation}
outlines the impact of this observation on our algorithm formulation,
but the gist of it is that with faster acting insulin we can remove
core abstractions, predictions, and simulation functionality typically
found in automated insulin delivery systems to simplify the \kernel.

Another impactful mechanism for simplicity is that ultimately, our
software impacts a biological system (the human injecting insulin),
which operates and longer timescales. Because we know we have O(hours)
until adverse effects take place, we can push some of our verification
tasks and correctness checks outside of the \kernel. Our closed-loop
replay app demonstrates this principle in action (Section
\ref{subsec:apps_redundancy}).

From an implementation perspective, we use Swift's Structured
Concurrency for simplified multithreading support, we use flat files
in the local file system to store serialized JSON objects rather than
Apple's CoreData SQL interface or HealthKit. We minimize caching,
opting to recompute values rather than managing caches. All in all,
these small implementation details add up to an overall system that is
easier for people to understand and reason about.

\subsection{Managing insulin pump and CGM hardware}
In \system we support one insulin pump: the Omnipod Dash. The \kernel
communicates directly with the Omnipod Dash via Bluetooth low energy,
and the Omnipod Dash supports three main commands for injecting
insulin. The first command injects a specified amount of insulin
effectively immediately. People use this command to inject a
calculated dose to cover meals typically. The second command sets a
low background rate of constant insulin injection. The physiological
basis for setting a rate is that the human body produces glucose
constantly to supply energy to the brain, muscles, liver, and anywhere
that needs it. This baseline rate covers the insulin needs for this
background glucose production. The third command temporarily overrides
the baseline rate with a specified new rate and duration.

In the \kernel, we change insulin by adjusting the baseline insulin
delivery rate temporarily each time we get a new CGM reading. Section
\ref{subsec:closed_loop_algorithm} describes our algorithm for making
these adjustments, but conceptually if the individual needs more
insulin, we increase the rate. If they need less, we decrease it.

By changing the baseline insulin delivery rate temporarily, we get
three desirable safety properties. First, if our software has a bug
and sets the baseline insulin delivery rate twice, it has no impact --
it is just a rate. In contrast, injecting twice delivers twice the
insulin, which we would like to avoid. Second, we configure the pump
with a maximum allowed rate, so that the pump hardware limits the
amount of insulin that the \kernel can inject, keeping it small enough
for the individual using it to correct easily by eating glucose if
needed. Third, if the pump becomes disconnected from the phone, the
temporary rate defaults back to the pre-specified baseline rate after
the temporary rate command expires, providing safety for the
individual.

Our general guidelines for pump configurations are to set the maximum
rate equal to 4x the baseline rate and always use a 30-minute duration
for our temporary rate commands. These settings enable us to move fast
enough to correct glycemic imbalance while ensuring that we limit the
amount of insulin that we can inject automatically. It is worth noting
that this general safety strategy comes from the OpenAPS
\cite{openaps} and Loop \cite{loop} open-source automated insulin
delivery systems, which we adopt in \system.

In \system, we support two CGM devices: the Libre 3 and Dexcom 7
CGMs. Both devices come with separate stand-alone apps that manage the
CGM, which we are in favor of from an architectural perspective. The
reason that we like this design is because these CGM apps have a slew
of functionality that they implement that is useful, like providing
parents with the ability to track their kid's glucose in real time, but
that we prefer to keep out of the \kernel.

The \kernel communicates with the Dexcom 7 using Bluetooth low energy
and communicates with the Libre 3 using a network service. Although
our current version supports both, soon we will drop support for the
Libre 3 in favor of the Dexcom 7 to avoid creating a dependency on a
network service and instead support only CGMs that we communicate with
directly using Bluetooth low energy.

\subsection{Event logs}
One interesting design decision for event logs is at what layer of
abstraction to log events. The tradeoff is lower in the stack (i.e.,
closer to the hardware), the more software state you can reproduce,
but the harder it is to work with the logs.

In our current implementation we log at the interface between the
LoopKit drivers and our core abstractions for facilitating closed loop
operation. This layer of abstraction is relatively high, meaning that
it is easy for apps to consume an infer the state, but we can only
reproduce high-level abstractions. In future work we will also log the
Bluetooth low energy messages to recreate the full software stack.

Our event logs hold events for the CGM, closed-loop algorithm runs,
and insulin pump events. CGM logs are the most straightforward
consisting only of a timestamp and glucose reading. Closed-loop
algorithm runs include the inputs to the closed-loop algorithm,
current therapeutic settings, and the result. The pump events include
all the commands that the \kernel issued to the pump and pump
alarms in case there is an issue that requires the individual's
attention. This data enables the apps that run on top of \system.

\subsection{Extending the \kernel with Loop}
Loop is another open-source automated insulin delivery system. Loop
provides functionality for helping calculate the amount of insulin
needed when one eats, tracking the absorption of that food
as it converts to glucose, and predicting how the individual's
metabolic state will evolve over the next six hours.

The interesting part about using Loop in \system is that it is an app
written by someone else that we ported to run within \system. The way
we support Loop is by creating virtual CGM and insulin pump devices
and then replaying our event logs to keep Loop's view of the state
consistent. This implementation for Loop allows us to use it for
calculating insulin doses, keeping track of carbohydrates, and
predicting future metabolic state while still using the \kernel to run
our own closed-loop algorithm in a separate and isolated protection
domain.

\subsection{Apps for redundancy and refinement}
\label{subsec:apps_redundancy}

The other three apps in \system, the Metabolic Watchdog (Section
\ref{subsec:metabolic_watchdog_algorithms}), settings analytics, and
closed-loop replay apps are all apps written by us to support the
overall functionality of \system, with the Metabolic Watchdog and
closed-loop replay apps providing redundancy to our system.

Architecturally, the closed-loop replay app provides an interesting
example of pushing complexity outside of the \kernel and taking
advantage of the longer timescales available when interacting with
biological systems. Our closed-loop replay app has a Python
implementation of our closed-loop algorithm. We run it and check the
results with the event logs to confirm that the closed-loop algorithm
is running as expected. In our implementation, we also include a slew
of sanity checks that make the source code a bit messy and add
complexity but are nice to have to confirm our assumptions. If this
app detects any inconsistencies, it notifies the individual and they
have plenty of time to adapt, if needed.

\subsection{Humans also provide redundancy}
In addition to all the automatic systems we have in place with
\system, the human that uses \system can also detect glycemic
imbalance or potentially dangerous situations. For example, the
individual who used \system, who we will call Bob, can sense both
hypoglycemia (low glucose, below 70 mg/dl) and hyperglycemia (high
glucose, above 200 mg/dl). When Bob's glucose goes low, he feels
simultaneously slightly drunk and panicky. When Bob's glucose goes
high, he gets a burning sensation in his feet. Since people living
with T1D deal with glycemic imbalance often, they can learn to feel
when it happens.

Additionally, Bob can feel insulin doses. When he injects insulin to
cover a meal, he gets a slight burning sensation at his pump insertion
site. Fortunately, he does not feel the background basal that his pump
injects. But by knowing when his pump is injecting insulin, he can
provide a backstop against the most egregious errors that can happen
with automated insulin delivery systems.

Our goal is to make sure that we avoid glycemic imbalance and
spurious injections, but if it does happen most people living with
T1D will be able to know about it.

\section{Background on T1D physiology}
\label{sec:background}

In this section, we outline the basics of human physiology for people
living with T1D. We cover food, digestion, and glucose, insulin,
insulin therapy for T1D, and important glucose levels. We introduce
these concepts to help provide background for our T1D management
strategy (Section \ref{sec:strategy}) and our closed-loop algorithmic
formulation (Section \ref{sec:algorithms}).

\subsection{Food, digestion, and glucose}
At a high-level, we can divide food into three macro nutrients:
carbohydrates, fat, and protein. These macro nutrients provide the body
with energy via glucose.

The main macro nutrient of interest to people living with T1D are
carbohydrates, or carbs. Carbs are sugar molecules and of the
macro nutrients have the largest impact on one's glucose. When people
eat carbs, they can take anywhere from 40 minutes to digest and
convert into glucose in your blood stream, or up to three hours
depending on what else you ate with them and how much your body needs
to process the carbs to turn them into glucose.

Fat and protein also convert to glucose, but at a much lower volume
per gram and much more slowly, taking five hours or more to fully
absorb.

In terms of absorption time, two extreme examples are pure glucose
tabs and pizza. Pure glucose tabs are already glucose, so the body can
absorb them with little digestive process. Pizza, on the other hand, is
high in both carbs through crust and in fat and protein through cheese
and toppings, resulting in total digestion time of upwards of five
hours more.

When we talk about glucose for T1D, we are referring to glucose
concentration, not raw amount of glucose. So 1g of carbs for a person
who weighs 200 lbs will raise their glucose by 3 mg/dl, whereas the
same amount of carbs for a person who weighs 100 lbs will result in a
rise of 5 mg/dl given the lower blood volume.

\subsection{Insulin semantics}
Strictly speaking, insulin in the bloodstream lowers glucose levels by
carrying glucose molecules from the blood to the liver, brain,
muscles, or stored as fat cells for later use.

In people with a healthy pancreas, their pancreas detects the
digestive process and secretes an appropriate amount of insulin
directly into the blood stream.

In contrast, people living with T1D inject insulin into the tissue
layer beneath the skin, so the insulin needs to travel from this sub
dermal layer to the bloodstream where it can attach to glucose
molecules, which takes longer than the pancreas.

Although the pancreas is good at matching insulin to glucose, it is
not perfect. Healthy people still experience a rise in glucose when
they eat high carb meals and can experience low glucose.

\subsection{Synthetic insulin activation}
The three main concepts for treating T1D using insulin are insulin on
board, insulin sensitivity, and insulin activation
curves. \emph{Insulin on board} defines the amount of injected insulin
that has \emph{not} yet attached to glucose molecules. This insulin
will activate over the next six hours and attach to a proportional
amount of glucose. The net effect of insulin attaching to glucose
molecules is lowering glucose levels.

\emph{Insulin sensitivity} defines how much of a glucose drop one
would expect if they took an international unit (or unit (U)) of
insulin. For example, if one's insulin sensitivity is 42 mg/dl / U,
then each unit of insulin injected in their body will reduce their
glucose by 42 mg/dl over the next six hours.

The rate at which insulin drops glucose during the six-hour absorption
period is defined by an exponential \emph{insulin activation
curve}. Typical insulin will take around 10 minutes to start acting,
peak anywhere between 55 minutes to 75 minutes depending on the type
of insulin and taper out over the final five or so hours of
activation. Insulin makers publish insulin activation curves for the
insulin they produce.

\subsection{Using insulin to manage T1D}
When using insulin to manage T1D, people need to consider both the
overall impact of the insulin as well as the dynamic response.

To cover a meal with insulin (\emph{bolus}), a typical insulin
treatment regime will require counting carbs for the meal, calculating
the amount of insulin needed overall, and then timing the insulin dose
to match carb absorption. For meals with fast acting carbs, people
will dose before they eat by 15 minutes or more, sometimes call
pre-bolusing. For slower absorbing carbs they dose when they eat. For
even slower absorbing meals (e.g., pizza), they will commonly split
their insulin dose into two or more bolus injections to match the
timing of their meal.

Counting carbs is hard and sometimes people living with T1D get it
wrong. If they dose too much insulin, it results in hypoglycemia (low
glucose) and they need to eat more carbs to correct it. If they dose
too little, they need to provide a \emph{correction}, which is an
additional insulin dose to bring their glucose back into range.

In addition to covering meals and correction doses, people living with
T1D also need to supply a baseline of effectively constant insulin
(\emph{basal insulin} or basal) to account for the glucose that one's
body produces naturally. Any insulin dosing to cover meals or provide
corrections are in addition to this baseline basal rate.

To help manage their T1D, people's doctors typically provide
therapeutic settings to help guide their calculations. These settings
include their basal rate, carb-to-insulin ratio, and insulin
sensitivity. With these settings, they use insulin to manage their
T1D.

\subsection{Important glucose levels}

\begin{figure}[t]
\centering
\includegraphics[width=\columnwidth]{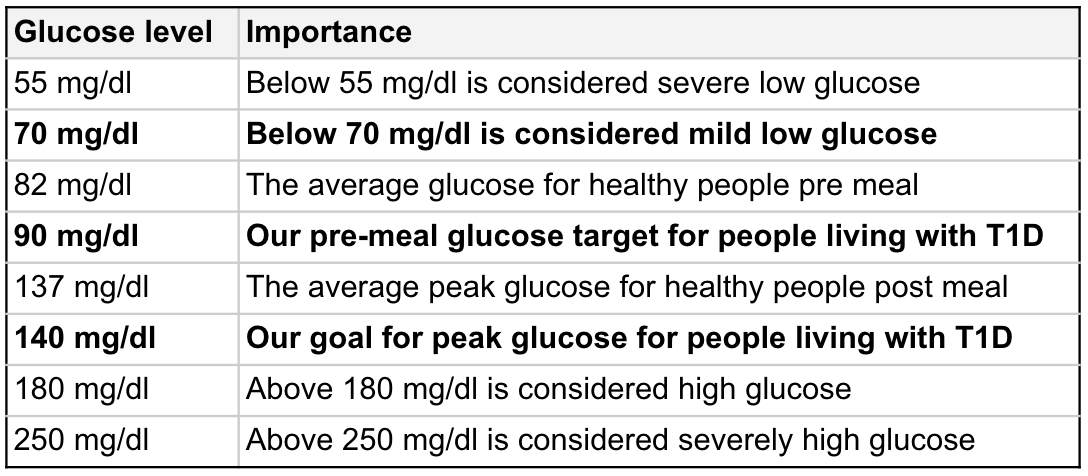}
\caption{Important glucose levels.}
\label{fig:important_glucose}
\hrulefill
\end{figure}

Figure \ref{fig:important_glucose} outlines meaningful glucose levels
with a description of their importance. We show levels for healthy
individual as well as the targets we set for \system. We differentiate
between glucose levels before someone eats (pre-meal) and for the 2-3
hours after they eat (post-meal).

\section{T1D management strategy}
\label{sec:strategy}

This section outlines the fundamentals of our T1D management
strategy. At its core our strategy is biohacking. We combine, diet,
physical activity, and automated insulin therapy to formulate our
strategy for managing T1D.

\subsection{Tight glucose management is key}
Our core hypothesis is that the key to avoiding health complications
due to T1D is tight glucose management.  However, our core hypothesis
is shaky. People living with T1D had their organs attacked by their
own immune system and most or all of the beta cells in their pancreas
are dead. These facts of T1D will almost certainly have a direct
impact on long-term health. There is nothing that people living with
T1D can do about these facts.

People living with T1D \emph{do} have control over their glucose
levels and there is evidence that elevated glucose levels are
connected with both negative short-term and long-term health effects
\cite{diabetes1993effect} \cite{genuth2006insights}. Informally, once
Bob started taking insulin when he was first diagnosed, he immediately
felt better -- he had more energy, was less cranky, he slept better --
there is merit to maintaining tight glucose control. Our hypothesis is
that people living with T1D can avoid negative long-term health
outcomes, and reverse them if they have already started, with tight
control over their glucose levels.

\subsection{Glucose targets for tight control}
Given that research has shown that elevated glucose levels increase
risk of long-term complications, we first need to define normal
glucose levels to set our targets. From a study done in 2007,
researchers found that young, healthy, and lean people who wore CGMs
had an average nighttime glucose level of 82 mg/dl with post-meal
spikes peaking at 137 mg/dl on average after eating high
glycemic-index foods \cite{freckmann2007continuous}. 

Given that healthy, young, and lean people define the category of
minimal risk for long-term health complications, we use their glucose
levels as guidance for our target glucose range. However, at least 42
factors impact glucose levels \cite{fourtytwofactors}, making
management with CGMs and injected insulin challenging. Thus, we
provide a bit of wiggle room for our pre-meal glucose range by setting
our target at 90 mg/dl, while staying between 70 mg/dl and 120 mg/dl
before meals. After eating, we try to keep our glucose levels between
70 mg/dl and 140 mg/dl.

\subsection{Management examples and tradeoffs}
To illustrate different management strategies and their tradeoffs, we
provide three example strategies and cast them as a classic ``pick two
of three'' problem.

The three goals that people living with T1D have are:
\begin{itemize}
  \item Minimize long-term health complications due to T1D with tight glucose control.
  \item Minimize the cognitive load needed to manage their metabolic state.
  \item Minimize dietary restrictions.
\end{itemize}

Our example strategies show how one can pick two of the three goals,
but trade off the third. In practice there are a number of points in
the overall management spectrum, but we do believe that making
tradeoffs is fundamental.

\begin{figure}[t]
\centering
\includegraphics[width=\columnwidth]{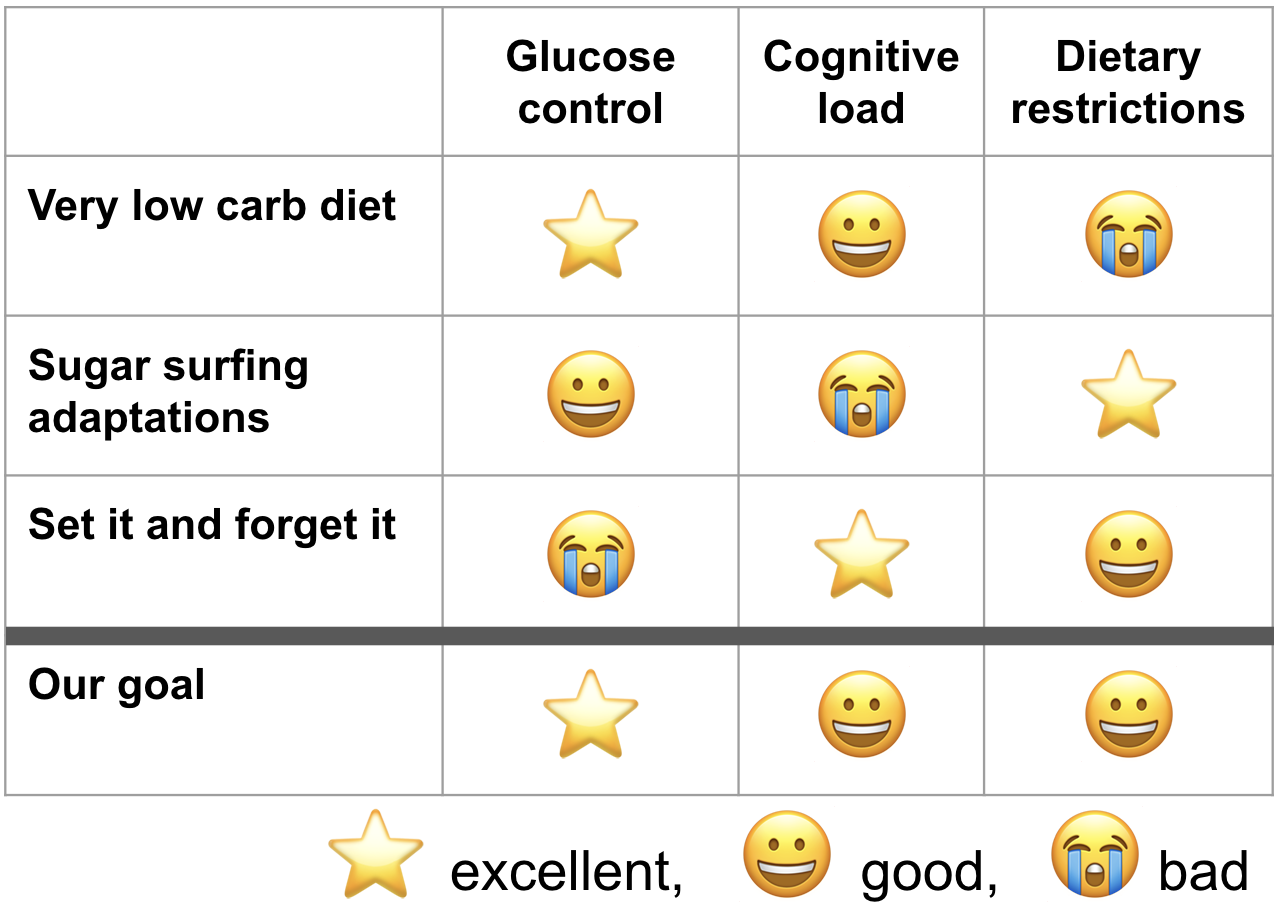}
\caption{Three example T1D management strategies and the tradeoffs
  they make and our goal for our management strategy.}
\label{fig:management_strategies}
\hrulefill
\end{figure}

Figure \ref{fig:management_strategies} shows three management
strategies and their tradeoffs. The first strategy is to eat a very
low carb diet \cite{lennerz2018management}. The intuition behind this
strategy is that eating fewer carbs requires less insulin. With fewer
carbs and less insulin, individuals will experience smaller swings in
glucose levels. With smaller swings in glucose levels, glycemic
control is easier. Although this strategy provides the best glycemic
control according to research, it brings with it substantial dietary
restrictions, where effectively people using this strategy are on a
Keto diet.

The second strategy is to constantly monitor CGM values and adapt to
accommodate likely glycemic imbalance before it happens
\cite{ponder2019sugar}. This basic strategy was introduced by Ponder
and McMahon in their book titled ``Sugar Surfing''
\cite{ponder2015sugar}. With Sugar Surfing, people can eat whatever
they want as long as they monitor their metabolic state and adapt by
eating glucose, taking a brisk walk, or injecting more insulin, to
account for how their body reacts. Although this strategy does away
with dietary restrictions, it brings with it a substantial cognitive
load from constantly monitoring glucose levels and reacting when
needed.

The third strategy is to use an automated insulin delivery system and
mostly ignore T1D. This basic strategy is epitomized by advances
introduced by Beta Bionics \cite{beta_bionics}, where their iLet
automated insulin delivery system automatically learns one's insulin
dynamics and injects insulin accordingly. In their clinical trial, the
iLet system showed that it could make profound improvements to people
with poor glycemic control. However, they were unable to improve
individuals who had moderate control already. In the end, this
management strategy does the most to reduce the cognitive load of
managing T1D, but is unable to achieve the tight glucose control that
we believe we need for minimizing long-term health complications.

\subsection{Our approach}
People will change their management strategies over time. For example,
Bob used variants of all three of the example strategies we outline
here and took with him the parts he liked while leaving behind the
parts he disliked. And since people living with T1D are biohackers at
their core, we expect their strategy to evolve with changes in their
goals and as they learn more about their individual physiology.

Given our anticipation for change, rather than outlining a
prescriptive management plan, we define the principles behind our
basic strategy so that people can pick and choose which principles
make the most sense for them.

\vspace{\baselineskip}
\begin{center}
  \begin{tabular}{ | p{0.95\columnwidth} | }
    \hline
    \textit{Principle: Eat carbs with a low glycemic index} \\
    \hline
  \end{tabular}
\end{center}
\vspace{\baselineskip}

In our experience, eating foods with a low \emph{glycemic index} is
important for maintaining glycemic balance. Glycemic index measures
how quickly carbs convert to glucose in the bloodstream, and eating
foods with a low glycemic index are easier to manage, especially when
using automated insulin delivery systems. For example, through
experimentation Bob found that he can eat beans but not bread, farro
but not rice, corn tortillas but not flour tortillas, and sweet
potatoes but not russet potatoes. By eating low glycemic-index foods
people can eat a medium carb diet while still maintaining tight
control over their glucose.

\vspace{\baselineskip}
\begin{center}
  \begin{tabular}{ | p{0.95\columnwidth} | }
    \hline \textit{Principle: Eat the same foods consistently and adapt
      insulin intake based on the past} \\ \hline
  \end{tabular}
\end{center}
\vspace{\baselineskip}

One of our main management strategy goals is to minimize the cognitive
load of managing T1D, and the time when the cognitive load is the
highest is when eating. When eating, people living with T1D need to
figure out what to eat, how much insulin to take, and when to take
it. Eating a moderate amount of carbs and avoiding high glycemic-index
foods helps with timing, so it really boils down to deciding what to
eat and how much insulin to take. For both of these decisions we have
found that consistency is key to reducing the cognitive load while
maintaining tight glucose control.

We recommend eating the same 20 or so foods over and over again and
using biohacking software to make it easier to calculate insulin
dosing based on the past. At mealtime, review the past few times you
ate the same meal and look at insulin dosing and the resulting glucose
after the meal to figure out how much insulin to take and the
timing. If your glucose stayed within your targets, take the same
amount of insulin. If you had a glucose spike or hypoglycemia, adjust
accordingly.

\vspace{\baselineskip}
\begin{center}
  \begin{tabular}{ | p{0.95\columnwidth} | }
    \hline \textit{Principle: Exercise vigorously every day to reduce
      insulin needs} \\ \hline
  \end{tabular}
\end{center}
\vspace{\baselineskip}

Exercise requires muscle contractions, which takes glucose, thus
lowering your overall insulin needs. The effects of exercise are both
immediate, as you are working out, and also carry over post exercise
for as many as 48 hours as your body refills its glycogen stores. It
adds some unpredictability while working out, so we wrote an app that
tracks standard exercise stats \emph{and} glucose in real time to help
people adapt while working out.

The net effect of exercise is increasing insulin
sensitivity. Increasing insulin sensitivity comes with lower insulin
dosing needs, which helps make it easier to maintain glycemic balance,
especially when using an automated insulin delivery system.

\vspace{\baselineskip}
\begin{center}
  \begin{tabular}{ | p{0.95\columnwidth} | }
    \hline \textit{Principle: Technology amplifies the effectiveness
      of biohacking} \\ \hline
  \end{tabular}
\end{center}
\vspace{\baselineskip}

The main overarching theme in all of our principles is using
technology to amplify our efforts. Whether it is from an app for
recording data after eating or an Apple Watch app for showing glucose
while working out, technology and biohacking software help make it
easier to manage T1D.

The most important technology is automated insulin delivery
systems. These closed-loop systems help with adaptations. The human
body is complex, and one's response to food can change from one day to
the next in ways that are difficult to anticipate. Having an automated
insulin delivery system adjust to these differences helps maintain
tight control with a lower cognitive load.

\subsection{Tying it together with an example}

\begin{figure}[t]
\centering
\includegraphics[width=\columnwidth]{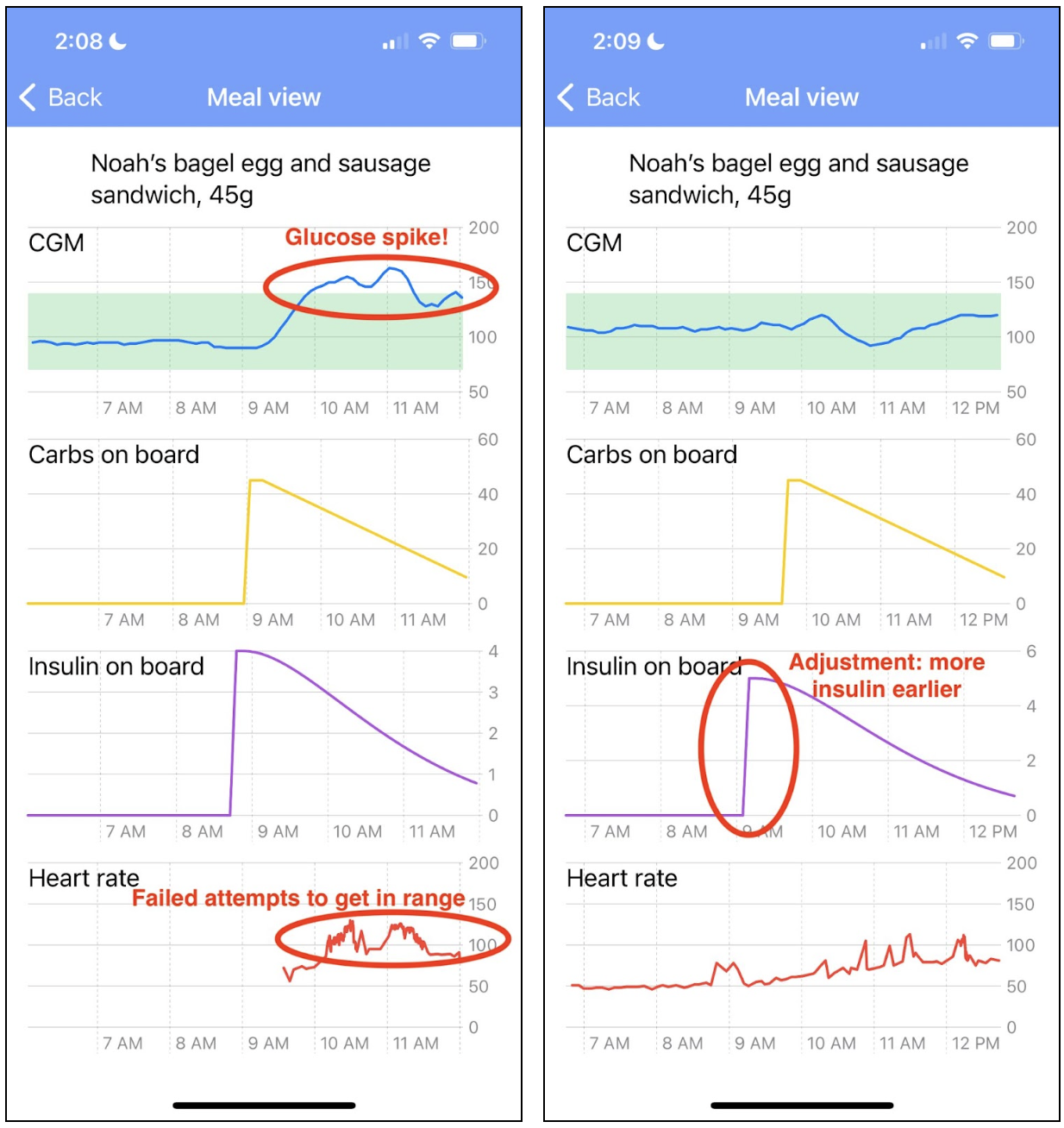}
\caption{An example of applying our principles in practice.}
\label{fig:noahs_example}
\hrulefill
\end{figure}

Figure \ref{fig:noahs_example} shows one example where Bob ate high
glycemic-index foods and used our software to adapt. This figure shows
two separate times that Bob ate a bagel sandwich from Noah's
Bagels. The first time he ate this meal (screen shot on the left) he
took a four-unit bolus ten minutes before he ate. When his glucose
levels went above his post-meal target range, he took two 25-minute
walks and still was unable to bring his glucose levels back down to
his pre-meal upper bound of 120 mg/dl. The next time he ate this meal
(screen shot on the right) he took a five-unit bolus and gave himself
40 minutes before eating. With this adjustment, he both avoided the
post-meal spike and kept his glucose in range as his meal digested.

This example highlights the principles in our approach. First, it
shows why we recommend avoiding high glycemic-index foods that are
high in carbs. When you do, typical adjustments are less
effective. Second, it shows that if you do want to eat high
glycemic-index foods that are high in carbs, eating the same meals
consistently and using past data helps to adjust dosing the next time
around. Third, technology is at the center of this entire process.

\section{Experiences}
\label{sec:algorithms}

One of the biggest challenges in managing T1D is the variability in
people's metabolic response to life. Differences in exercise, timing
of insulin and eating, macro nutrient combinations, stress, caffeine,
and so on all have an unpredictable impact on people's glucose levels.

To cope with these differences, closed-loop automated insulin delivery
systems monitor glucose levels in real time and adjust insulin to
compensate automatically. If an individual's glucose level is likely
to go low, the system will shut off insulin to bring their glucose
levels back up using their body's background glucose production. If
their glucose levels are likely to go high, it will inject
insulin. These automatic adjustments are what make closed-loop systems
so effective at managing T1D.

In this section, we provide a qualitative description of our
experiences designing, implementing, and running \system. We cover our
core algorithm development and iterations, and report on our
experiences using iOS for a security-critical app. Section
\ref{sec:eval} describes our quantitative evaluation.

Our overarching principle is that we focus on usefulness to the human
using \system. At our core, the actions we take are specifically for
the human who is living with T1D. We start from problems \emph{they}
have and design systems to solve these problems.

We have been working with one individual, who we will call Bob, for
six months as we design and implement this system. Bob used the
Metabolic Watchdog for the entire six months and used our closed-loop
system for one week. Based on Bob's feedback and results, we iterate
on our design.

\subsection{Metabolic Watchdog algorithms}
\label{subsec:metabolic_watchdog_algorithms}

\begin{figure}[t]
\centering
\includegraphics[width=0.75\columnwidth]{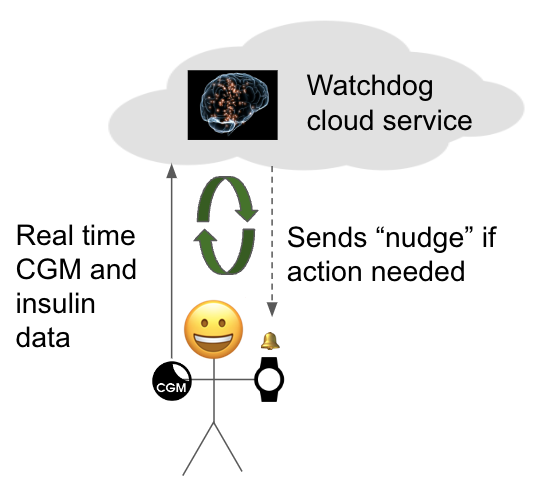}
\caption{Overview of the Metabolic Watchdog.}
\label{fig:watchdog_overview}
\hrulefill
\end{figure}

Our high-level goal for our Metabolic Watchdog is to send a nudge to
people before they experience glycemic imbalance automatically (Figure
\ref{fig:watchdog_overview}). Glycemic imbalance is when a person's
glucose level drops below 70 mg/dl or rises above 180 mg/dl. When the
algorithm detects likely glycemic imbalance, it sends a nudge using a
push notification we send to the individual's Apple Watch and
iPhone. Conceptually, these nudges play the same role that a
closed-loop algorithm plays in an automated insulin delivery system,
but for people running open-loop (e.g., manual injections using a
syringe).

To maintain simplicity for the \kernel, we run the Metabolic Watchdog
as a separate app, outside of the core \kernel.

\begin{figure}[t]
\centering
\includegraphics[width=0.75\columnwidth]{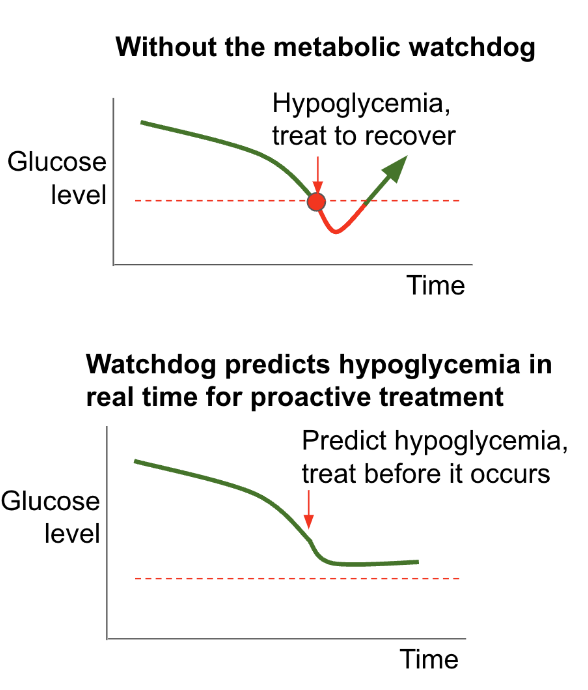}
\caption{Preventing low glucose with the Metabolic Watchdog.}
\label{fig:watchdog_operation}
\hrulefill
\end{figure}

The ideal nudge is both timely and actionable so they can correct the
likely imbalance before it happens and only get alerts when there is
an action they can take to correct it. Figure
\ref{fig:watchdog_operation} shows the difference in glucose levels
for an individual experiencing low glucose levels without the
Metabolic Watchdog and using the Metabolic Watchdog. With traditional
level-based alerting, the individual would be notified once they have
already experienced low glucose levels, but with the Metabolic
Watchdog they receive a notification on their Apple Watch indicating
that they are likely to experience hypoglycemia without intervention,
providing them with enough time to correct before their glucose level
goes low.

Algorithmically, glucose prediction has been well studied in the
literature \cite{arora2023multivariate}. Our contribution is showing
how simple prediction models are effective in practice, obviating the
need for more complex predictions, and how we need to consider insulin
on board in our formulation.

To predict glycemic imbalance, we use a linear regression model that
looks 15 minutes into the future. If we detect glycemic imbalance in
the future, we nudge the individual.

This simple prediction for low glucose worked, but predictions for
high glucose took a few iterations. We noticed that Bob would
sometimes experience low glucose after treating for high glucose by
taking a brisk walk. The root cause was that Bob had enough insulin
on board to compensate for his elevated glucose levels. Updating
our high prediction algorithm to account for insulin on board in
addition to our glucose prediction eliminated these false positive
alerts.

One surprise with the prediction for low glucose was that our simple
model was effective. In practice, Bob effectively never faced a false
negative (missing low glucose) when using our model. We had
anticipated needing to use a more sophisticated model and using the
simple model as a baseline. But, after getting real-world experience
with the simple model, we found that it was good enough to solve Bob's
problem of detecting low glucose before it happened.

The Metabolic Watchdog also helped improve Bob's overall glycemic
control, which we detail in our evaluation (Section \ref{sec:eval}).

\subsection{Closed-loop algorithm motivation}
\label{subsec:closed_loop_algorithm_motivation}

One recent trend that we observe is the advent of ultra-fast acting
insulin. Over the last few years Lyumjev insulin and Fiasp insulin
both provide the ability for insulin to go from injection to the blood
stream up to 30\% faster than more traditional fast-acting insulin
\cite{leohr2023ultra}.

With ultra-fast insulin action, it reduces the need for our algorithms
to predict the future since it can simply react rather than
predict. One concrete impact of this difference is that the \kernel
does \emph{not} have an abstraction for carbohydrates. When carbs
digest, the body converts food into glucose in the blood stream, and
most or all other automated insulin delivery systems predict
carbohydrate absorption to compensate. But given that we start with
ultra-fast insulin, we choose to react to rising glucose levels rather
than predict, which simplifies our system because we avoid introducing
an additional abstraction and avoid designing prediction algorithms in
the \kernel.

\subsection{Closed-loop algorithm}
\label{subsec:closed_loop_algorithm}

In general, closed-loop algorithms work by reading the latest glucose
value from the CGM every five minutes, calculating the difference
between current glucose and a target glucose level, and adjusting
insulin to reach the target.

To calculate the amount of insulin needed, we use a combination of the
amount of insulin on board with the insulin sensitivity to determine
the amount of insulin needed or the insulin surplus. Based on this
calculation, we adjust the basal rate to compensate.

From the user perspective, we only ask the user to set two therapeutic
settings: their minimum basal rate and their insulin
sensitivity. Using these two settings we design algorithms that can
compensate for changes in practice with these settings.

From a theoretical perspective, our algorithm for this closed-loop
problem is a proportional controller from control theory (the ``P'' in
a PID controller). The advantage of using control theory is that we
can use the principled tools available for feedback control to analyze
our system and reason about algorithmic tradeoffs mathematically.

From a practical perspective, our real-world deployment has uncovered
several challenges that we address. First, we include a safety
threshold where anytime the individual's glucose drops below 80 mg/dl,
we shut off insulin delivery. The transitions between the shut off
state and closed loop control need to be handled carefully because
the insulin on board values will continue to decrease while the system
suspends insulin delivery, so when the closed loop algorithms turn
back on the state of the system is different\footnote{In control
theory, there is a similar concept called \emph{integrator windup},
which happens from non-linear behavior in the system.}. Second, basal
rates change throughout the day, and by using a proportional controller
the system will reach a steady state equilibrium with a persistent
error, which we observed with Bob. In other words, we can set our
target glucose level for 90 mg/dl, but if the actual basal rate is
lower than what we have configured, it will stabilize at a lower
steady-state value.

We are in the process of adding an additional controller that
accumulates values as persistent errors happen (i.e., the ``I'' of a
PID controller) to compensate, but we are still in the design phase
for this new controller.

Using feedback control for closed-loop automated insulin delivery has
been well studied in the literature \cite{shi2019feedback}, mostly in
simulation or mathematically. Our contribution is using a simple
algorithm that has the properties we care about for building a real
closed-loop system that facilitates tight control over glucose levels
for people living with T1D.

\subsection{Experience using iOS for \system}
\label{subsec:ios_experience}

Overall, we are happy with iOS as a platform for running \system. The
development tools are solid, the libraries are easy to use and provide
clean abstractions, and the overall security model for iOS, with strong
isolation between apps, is a match for our overall design.

There are three main areas, where based on our experiences, could be
improved to support this type of system. First, we use Apple's
Structured Concurrency abstractions for synchronization. Conceptually,
using Structured Concurrency is similar to using a Mesa monitor with
compiler support to eliminate data races. We generally liked using
Structured Concurrency and feel that it helped us simplify our \kernel
implementation. However, the one aspect that we believe might be
problematic is that, like monitors, when an actor object makes an
async call, it gives up the lock and the internal state of the object
can change before the caller reacquires the lock and continues
running. These semantics mean that people need to ensure that their
internal object state is consistent before calling an async routine
and they need to recheck any assumptions when it returns because the
object's state can change. For experts who are experienced using
monitors this type of reasoning is appropriate, but it might be
difficult for people who are less experienced using monitors.

Second, Apple's support for background execution doesn't support
periodic tasks well. Our first version of the Metabolic Watchdog ran
completely on device, using background execution to wake up every five
minutes and run predictions. In practice, these periodic tasks can
take an hour or more for the next invocation. For Bob, this meant that
in our first implementation there were a few cases where alerts should
have fired but did not due to the scheduling policy. These missing
alerts caused Bob to lose trust in the system, so we moved to a
cloud-based implementation. We believe that there are abstractions
that iOS could support to facilitate this style of computation, but
the current abstractions are not suitable. In contrast, the \kernel
does have reliable run time invocation because it uses the external
Bluetooth devices to notify iOS to run tasks in the background for the
\kernel.

Third, Bluetooth access control is too broad, which has both
advantages and disadvantages. In iOS, apps request broad Bluetooth
permissions, which gives them access to all Bluetooth devices. This
level of access is an advantage because we can access the CGM at the
same time as the CGM manufacturer's app, which is a fundamental
requirement for running an automated insulin delivery system. However,
this also means that any apps with Bluetooth access can access the
insulin pump, which is problematic. Ideally, we would like to have
app-level permissions for certain classes of devices, like CGMs and
insulin pumps.

\begin{figure*}[t]
\centering
\includegraphics[width=0.8\textwidth]{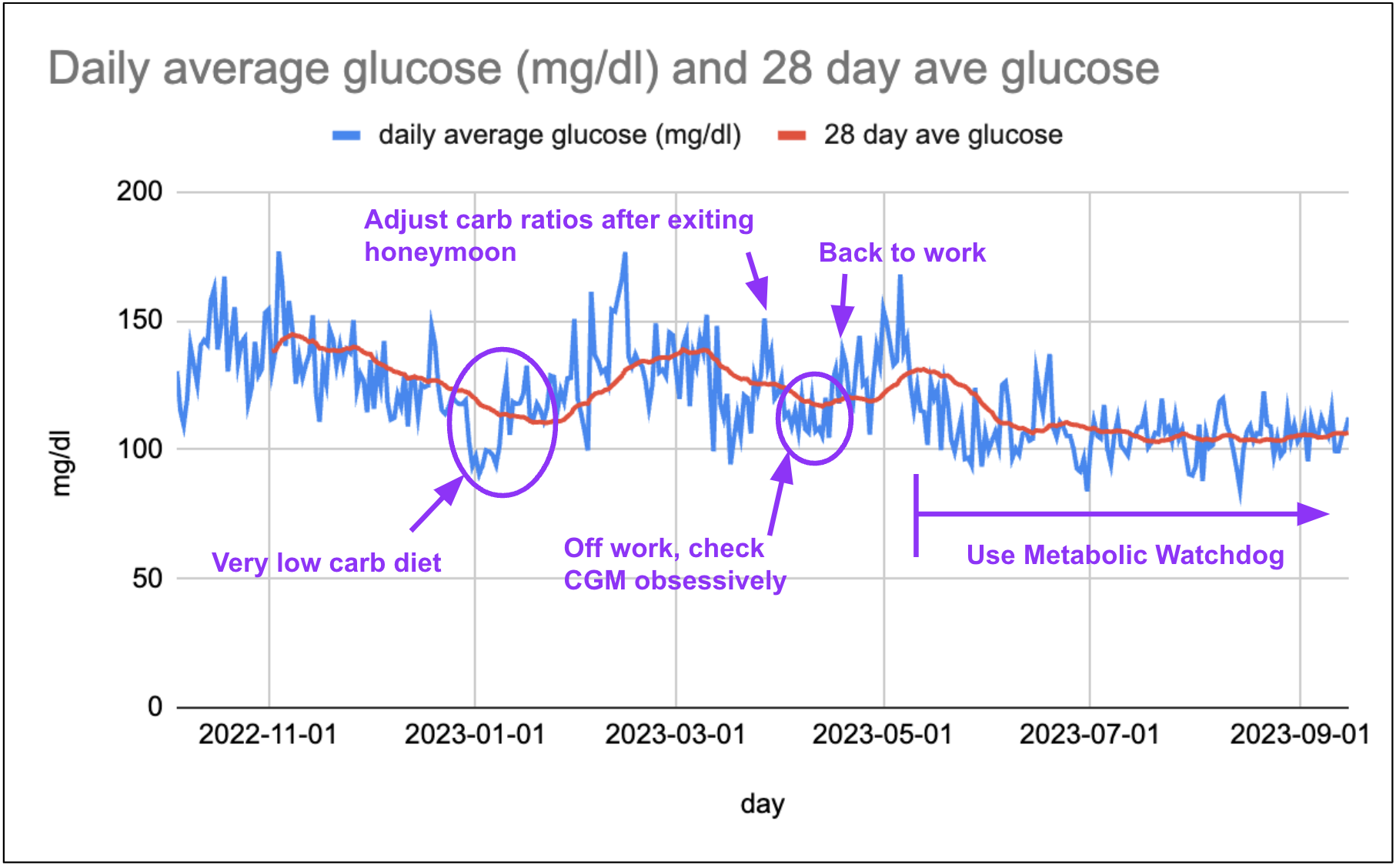}
\caption{Daily average glucose for Bob with annotations about life
  events and biohacking experiments. These results show tighter
  glycemic control once Bob started using the Metabolic Watchdog.}
\label{fig:annotated_glucose}
\hrulefill
\end{figure*}

\section{Evaluation}
\label{sec:eval}

In this section, we evaluate \system. Our core hypothesis is that we
can design a system for biohacking and managing T1D using security
first principles from the start, which has a simple implementation yet
still supports tight control over T1D.

For our evaluation, we use data from one individual, who we will call
Bob, who has been using \system for the last six months to manage his
T1D. We show results for Bob's average daily glucose, the primary
measure of glycemic control for people living with T1D, for six months
preceding his use of the Metabolic Watchdog and then for his first
four months using it to compare. We also show results for one week's
worth of data when Bob was running the \kernel to manage his T1D using
our closed-loop algorithms. We also compare the complexity of our
\kernel implementation against another open-source automated insulin
delivery system, Loop.

From a therapeutic goals perspective, Bob is striving to achieve
glycemic control that is in line with healthy individuals, which is
more aggressive than what the American Diabetes Society outlines. The
reason he wants to be aggressive is that when he was diagnosed in
2022, he had already incurred long-term damage from elevated glucose
levels before diagnosis. Now that he has a T1D diagnosis, he tries to
maintain glucose levels that are consistent with healthy people, while
minimizing the cognitive load needed to do so.

Bob is a person living with T1D and must inject insulin every day,
which is risky. He already takes on the inherent risk of putting his
life in the hands of medical devices and injecting dangerous
hormones. He is an enthusiastic biohacker and constantly running
experiments on himself. All of this is just Tuesday for Bob.

Bob was unwilling to use Loop, another automated insulin delivery
system. He felt that Loop was too complex and did not want to have
that level of complexity controlling his insulin pump. When we
introduced Bob to \system, he enthusiastically volunteered to run the
first version of our software ready for human use.

To run our experiments, we use \system running on Bob's iPhone 14. Bob
uses a Libre 3 CGM and a combination of manual insulin injections
using a syringe and Humalog insulin in addition to insulin injections
from a Omnipod Dash insulin pump with Lyumjev insulin. Bob also wears
an Apple Watch, which we use to deliver notifications from the
Metabolic Watchdog.

\subsection{Ethical considerations}
Since this study is an n=1 evaluation, and Bob is a willing and
enthusiastic participant, we do not need to go through the IRB
process. However, to ensure that we avoid putting Bob at
risk, we went over our plan in detail with his endocrinologist,
diabetes educator, and dietitian. Also, we got feedback from a sports
medicine doctor and an emergency room doctor. Plus, we asked Bob to go
over the plan with his therapist to ensure that these experiments
avoid mental health issues.

All in all, we understand the gravity of having humans inject
dangerous hormones using our software and have taken steps to consult
outside experts to ensure that what we are doing makes sense from a
medical perspective.

\subsection{Glycemic control with the Metabolic Watchdog}
\label{subsec:eval_metabolic_watchdog}

Figure \ref{fig:annotated_glucose} shows Bob's daily average glucose
level, the rolling 28-day average, and has annotations to highlight
key events during this time. Bob's first attempt at controlling
glucose levels closer to the non-diabetic range was when he tried
eating a very low carb diet (60g or less of carbs per day) in early
2023. Although his glucose results were what he was looking for, he
did not like restricting his diet so severely and anecdotally he felt
tired while eating so few carbs (he exercises a lot). People have
reported that the tired feeling goes away, but the dietary
restrictions alone were enough for him to cut that experiment short.

In mid-March 2023, Bob's pancreas stopped producing insulin, forcing
him to cover all meals and reduce his carb-to-insulin ratio drastically
(previously he could eat 20g-30g of carbs without needing to
bolus). When people are first diagnosed with T1D they often continue
to produce a small amount of insulin, during what is referred to as
the ``honeymoon period,'' which ended for Bob in March 2023.

During late March and early April, Bob took a few weeks off work to
spend some time learning about the fundamentals of diabetes
management, and during this time he looked at his CGM data
constantly. As a result, he was able to manage his glucose
effectively.

In April, Bob went back to work, and his control suffered as he was
unable to micromanage his glucose levels all day.

In early May, Bob started using the Metabolic Watchdog to help improve
his glycemic control.

\begin{figure}[t]
\centering
\includegraphics[width=\columnwidth]{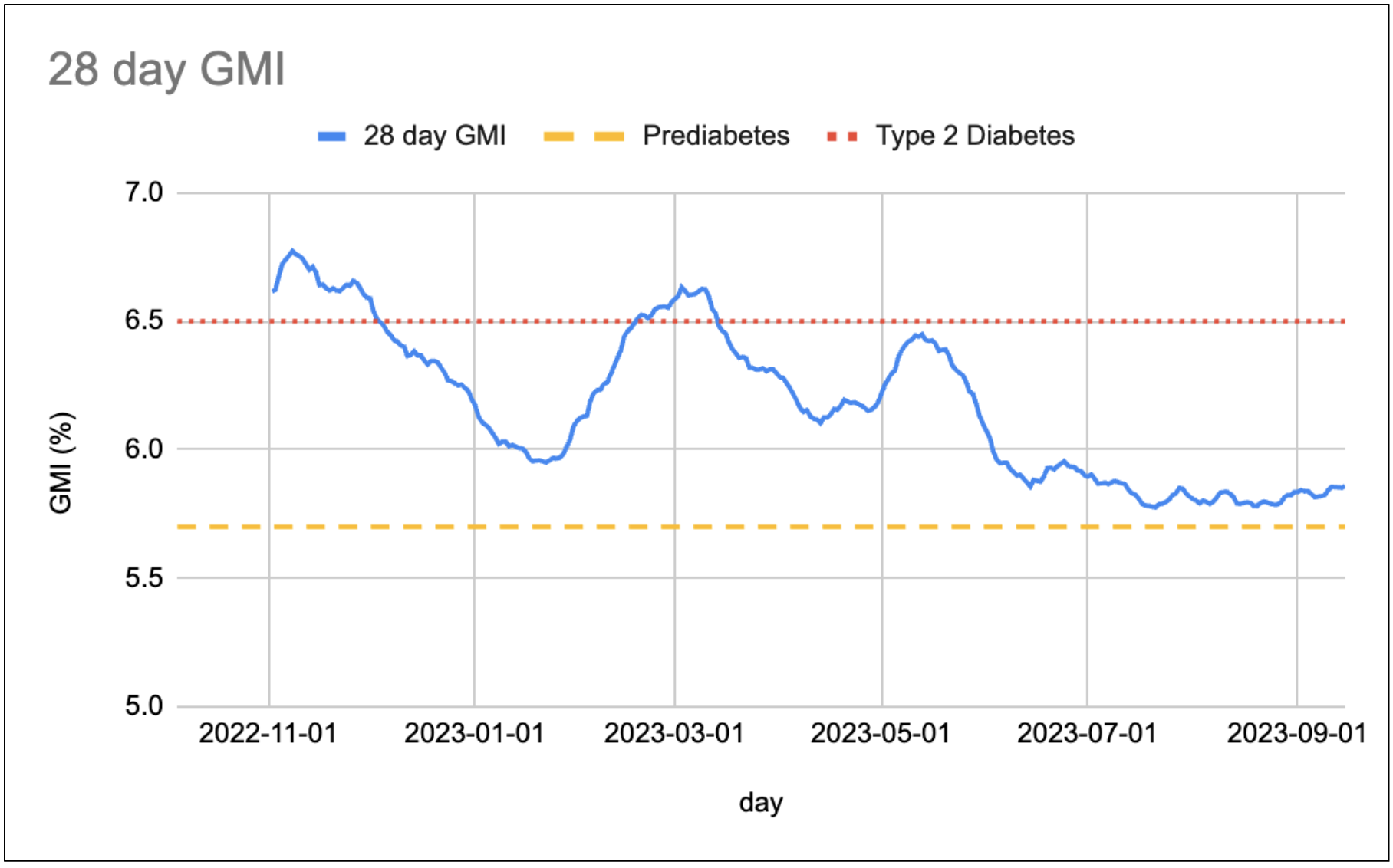}
\caption{GMI for Bob compared to GMI levels for Prediabetes and Type 2
  Diabetes. These results show that Bob has nearly the same average
  glucose level as a healthy individual, despite living with T1D.}
\label{fig:gmi_results}
\hrulefill
\end{figure}

Figure \ref{fig:gmi_results} shows Bob's GMI
\cite{bergenstal2018glucose}, which approximates A1C numbers based on
CGM data. It shows Bob's GMI for a 28-day rolling average of his
glucose levels. At its peak, his GMI was 6.8\% and after using the
Metabolic Watchdog in early May 2023 his GMI settled into the 5.7\% -
5.9\% range consistently. These average glucose levels are still in
the low end of the “Prediabetes” range for people who have insulin
resistance (i.e., Type 2 Diabetes), but at an acceptable level for Bob
as someone living with T1D. These glucose levels are
consistent with people who eat a very low carb diet, which is the most
effective management technique for T1D currently
\cite{lennerz2018management}.

Since using the Metabolic Watchdog, Bob traveled to the American South
(there was fried food everywhere), traveled internationally, worked in
a high-stress job, worked out a ton, and got COVID. Despite these
difficulties, Bob was able to maintain tight control for the entire
period with decreasing cognitive load as we refined our
software.

\subsection{Maintaining glycemic control with the \kernel closed-loop algorithm}

During one week in late November 2023 - early December 2023, Bob ran
the \kernel closed-loop algorithm. His goal was to have the \kernel
automate some of the adaptations he made when using the Metabolic
Watchdog alone. Thus, his goal for this experiment was to maintain the
same level of control that we demonstrate in Section
\ref{subsec:eval_metabolic_watchdog}.

Over this time, Bob had a GMI of 5.9\%, which is consistent
with the results obtained from our Metabolic Watchdog
study. Anecdotally, Bob reported having to make fewer adaptations as
the system adjusted insulin to accommodate his CGM readings. He
continued to use the Metabolic Watchdog in concert with the
closed-loop system.

\subsection{\kernel complexity}
\label{subsec:biokernel_complexity}

To evaluate the complexity of our \kernel, we count the lines of code
in our implementation and compare against Loop, another open-source
automated insulin delivery system.

Using the ``cloc'' utility the \kernel app has 4.6k lines of code
compared to 39.8k lines of code in the Loop app, an order of magnitude
reduction. We omit the lines of code coming from LoopKit and other
drivers because these are shared between both projects. However,
LoopKit and the drivers have a substantial amount of code, weighing in
at 70.1k lines of code and is likely the next big opportunity for
simplification.

All those extra lines of code in Loop are useful, which is why we
port Loop to run within \system. Bob uses it for meal announcements
and looking at their simulation and prediction results. However, for
the core closed-loop algorithm we show how to keep it separated in an
isolated protection domain while still providing the right interfaces
to enable a fully featured automated insulin delivery system.

To explain why we have such a large reduction in source
code, we outline the differences between the two systems. First, Loop
has several features that we move outside of the \kernel. These
features include a remote interface for insulin dosing (which we think
it a bad idea in general), a Watch app, Siri command interfaces, third
party libraries, tutorials, meal announcements, physiological
simulations, and predictions of future metabolic states. Second, we
simplify our implementation of features that are shared between the
\kernel and Loop. These shared features include a simplified
local storage implementation, simplified concurrency support, minimal
therapeutic settings, and a simplified closed loop algorithm. As our
evaluation shows, we can still provide tight control over T1D with
this simplified core system and still support most or all these
additional features architecturally.

\section{Related work}
In addition to the research that we already mention in this paper,
several other studies are also related to our work on \system.

Several automated insulin delivery systems exist today. With companies
such as Tandem, Insulet, Medtronic, and Beta Bionics all providing
closed loop systems that connect CGMs to insulin pumps for automatic
insulin delivery. From the open source world, OpenAPS \cite{openaps}
and Loop \cite{loop} also provide systems that people can use. Our
study builds on top of these works, where we use many of the safety
principles from OpenAPS, the software from Loop, and the push for
user-facing simplicity from Beta Bionics. However, our focus is on how
to decompose these monolithic systems into extensible and isolated
components.

Previous research has looked at the security of implanted medical
devices in general \cite{halperin2008security, burleson2012design,
  rushanan2014sok}, in addition to looking at insulin pumps in
particular \cite{li2011hijacking, paul2011review}, with more recent
work looking at providing improved security
\cite{marin2016feasibility, ahmad2018securing}. Also, recent work has
looked at applying formal methods to insulin pumps for high assurance
\cite{panda2021secure}. These works focus on the device and their
communication channel. In contrast, with \system we assume that these
devices are correct and secure and focus our efforts on the software
we use to run the automated insulin delivery system while providing
redundancy in case anything goes wrong.

In our \kernel, we expose functionality through Universal Links where
other apps can open UI views from the \kernel to enable the individual
to access the system. Work from Roesner et al
\cite{roesner2013securing} outlines how this can work securely on
Android devices, and Flexdroid \cite{seo2016flexdroid} shows how to
provide more fine-grained isolation within an Android app.

\section{Conclusion}
Our audacious long-term goal is to turn a T1D diagnosis from a death
sentence into an indicator of longevity, where people living with T1D
will be expected to live longer then their healthy peers. This
longevity will come by virtue of the tight control that they maintain
over their metabolism through biohacking and advanced computer
systems. Our first step towards this goal is to ensure that people can
use trustworthy computer systems to manage their glucose levels.

In this paper, we showed how a clean slate approach to designing and
implementing the new Metabolic Operating System led to a system
built with security principles from the start, yet maintained the
functionality needed to manage people's T1D. Our implementation was
simple, keeping a spartan \kernel app that managed the most critical
parts of our overall system, while it provided event logs for other
apps to consume so that they could contribute to the overall
management problem safely.

\bibliographystyle{plain}
\bibliography{typezero}

\begin{thebibliography}{10}

\bibitem{accetta1986mach}
Mike Accetta, Robert Baron, William Bolosky, David Golub, Richard Rashid,
  Avadis Tevanian, and Michael Young.
\newblock Mach: A new kernel foundation for unix development.
\newblock 1986.

\bibitem{ahmad2018securing}
Usman Ahmad, Hong Song, Awais Bilal, Shahzad Saleem, and Asad Ullah.
\newblock Securing insulin pump system using deep learning and gesture
  recognition.
\newblock In {\em 2018 17th IEEE International Conference On Trust, Security
  And Privacy In Computing And Communications/12th IEEE International
  Conference On Big Data Science And Engineering (TrustCom/BigDataSE)}, pages
  1716--1719. IEEE, 2018.

\bibitem{arora2023multivariate}
Sunny Arora, Shailender Kumar, and Pardeep Kumar.
\newblock Multivariate models of blood glucose prediction in type1 diabetes: A
  survey of the state-of-the-art.
\newblock {\em Current Pharmaceutical Biotechnology}, 24(4):532--552, 2023.

\bibitem{baca2007tear}
Justin~T Baca, David~N Finegold, and Sanford~A Asher.
\newblock Tear glucose analysis for the noninvasive detection and monitoring of
  diabetes mellitus.
\newblock {\em The ocular surface}, 5(4):280--293, 2007.

\bibitem{benoit2018trends}
Stephen~R Benoit, Yan Zhang, Linda~S Geiss, Edward~W Gregg, and Ann Albright.
\newblock Trends in diabetic ketoacidosis hospitalizations and in-hospital
  mortality—united states, 2000--2014.
\newblock {\em Morbidity and Mortality Weekly Report}, 67(12):362, 2018.

\bibitem{bequette2005critical}
B~Wayne Bequette.
\newblock A critical assessment of algorithms and challenges in the development
  of a closed-loop artificial pancreas.
\newblock {\em Diabetes technology \& therapeutics}, 7(1):28--47, 2005.

\bibitem{bergenstal2018glucose}
Richard~M Bergenstal, Roy~W Beck, Kelly~L Close, George Grunberger, David~B
  Sacks, Aaron Kowalski, Adam~S Brown, Lutz Heinemann, Grazia Aleppo, Donna~B
  Ryan, et~al.
\newblock Glucose management indicator (gmi): a new term for estimating a1c
  from continuous glucose monitoring.
\newblock {\em Diabetes care}, 41(11):2275--2280, 2018.

\bibitem{fourtytwofactors}
Adam Brown.
\newblock 42 factors that affect blood glucose?! a surprising update, 2022.
\newblock
  \url{https://diatribe.org/42-factors-affect-blood-glucose-surprising-update}.

\bibitem{brown2021multicenter}
Sue~A Brown, Gregory~P Forlenza, Bruce~W Bode, Jordan~E Pinsker, Carol~J Levy,
  Amy~B Criego, David~W Hansen, Irl~B Hirsch, Anders~L Carlson, Richard~M
  Bergenstal, et~al.
\newblock Multicenter trial of a tubeless, on-body automated insulin delivery
  system with customizable glycemic targets in pediatric and adult participants
  with type 1 diabetes.
\newblock {\em Diabetes Care}, 44(7):1630--1640, 2021.

\bibitem{burleson2012design}
Wayne Burleson, Shane~S Clark, Benjamin Ransford, and Kevin Fu.
\newblock Design challenges for secure implantable medical devices.
\newblock In {\em Proceedings of the 49th annual design automation conference},
  pages 12--17, 2012.

\bibitem{beta_bionics}
Luz~E. Castellanos, Courtney~A. Balliro, Jordan~S. Sherwood, Rabab Jafri,
  Mallory~A. Hillard, Evelyn Greaux, Rajendranath Selagamsetty, Hui Zheng,
  Firas~H. El-Khatib, Edward~R. Damiano, and Steven~J. Russell.
\newblock {Performance of the Insulin-Only iLet Bionic Pancreas and the
  Bihormonal iLet Using Dasiglucagon in Adults With Type 1 Diabetes in a
  Home-Use Setting}.
\newblock {\em Diabetes Care}, 44(6):e118--e120, 06 2021.

\bibitem{cobelli2011artificial}
Claudio Cobelli, Eric Renard, and Boris Kovatchev.
\newblock Artificial pancreas: past, present, future.
\newblock {\em Diabetes}, 60(11):2672--2682, 2011.

\bibitem{diabetes1993effect}
Diabetes Control and Complications Trial~Research Group.
\newblock The effect of intensive treatment of diabetes on the development and
  progression of long-term complications in insulin-dependent diabetes
  mellitus.
\newblock {\em New England journal of medicine}, 329(14):977--986, 1993.

\bibitem{cryer2012severe}
Philip~E Cryer.
\newblock Severe hypoglycemia predicts mortality in diabetes.
\newblock {\em Diabetes care}, 35(9):1814--1816, 2012.

\bibitem{defronzo2015type}
Ralph~A DeFronzo, Ele Ferrannini, Leif Groop, Robert~R Henry, William~H Herman,
  Jens~Juul Holst, Frank~B Hu, C~Ronald Kahn, Itamar Raz, Gerald~I Shulman,
  et~al.
\newblock Type 2 diabetes mellitus.
\newblock {\em Nature reviews Disease primers}, 1(1):1--22, 2015.

\bibitem{cdc_type2}
Centers for Desease~Control and Prevention.
\newblock National diabetes statistics report, 2023.
\newblock \url{https://www.cdc.gov/diabetes/data/statistics-report/index.html}.

\bibitem{freckmann2007continuous}
Guido Freckmann, Sven Hagenlocher, Annette Baumstark, Nina Jendrike, Ralph~C
  Gillen, Katja R{\"o}ssner, and Cornelia Haug.
\newblock Continuous glucose profiles in healthy subjects under everyday life
  conditions and after different meals.
\newblock {\em Journal of diabetes science and technology}, 1(5):695--703,
  2007.

\bibitem{genuth2006insights}
Saul Genuth.
\newblock Insights from the diabetes control and complications
  trial/epidemiology of diabetes interventions and complications study on the
  use of intensive glycemic treatment to reduce the risk of complications of
  type 1 diabetes.
\newblock {\em Endocrine Practice}, 12:34--41, 2006.

\bibitem{t1d_numbers}
Gabriel~A Gregory, Thomas I~G Robinson, Sarah~E Linklater, Fei Wang, Stephen
  Colagiuri, Carine {de Beaufort}, Kim~C Donaghue, Jessica~L Harding, Pandora~L
  Wander, Xinge Zhang, Xia Li, Suvi Karuranga, Hongzhi Chen, Hong Sun, Yuting
  Xie, Richard Oram, Dianna~J Magliano, Zhiguang Zhou, Alicia~J Jenkins,
  Ronald~CW Ma, Dianna~J Magliano, Jayanthi Maniam, Trevor~J Orchard, Priyanka
  Rai, and Graham~D Ogle.
\newblock Global incidence, prevalence, and mortality of type 1 diabetes in
  2021 with projection to 2040: a modelling study.
\newblock {\em The Lancet Diabetes \& Endocrinology}, 10(10):741--760, 2022.

\bibitem{grier2008secure}
Chris Grier, Shuo Tang, and Samuel~T King.
\newblock Secure web browsing with the op web browser.
\newblock In {\em 2008 IEEE Symposium on Security and Privacy (sp 2008)}, pages
  402--416. IEEE, 2008.

\bibitem{halperin2008security}
Daniel Halperin, Thomas~S Heydt-Benjamin, Kevin Fu, Tadayoshi Kohno, and
  William~H Maisel.
\newblock Security and privacy for implantable medical devices.
\newblock {\em IEEE pervasive computing}, 7(1):30--39, 2008.

\bibitem{hartig1997performance}
Hermann H{\"a}rtig, Michael Hohmuth, Jochen Liedtke, Sebastian Sch{\"o}nberg,
  and Jean Wolter.
\newblock The performance of $\mu$-kernel-based systems.
\newblock {\em ACM SIGOPS Operating Systems Review}, 31(5):66--77, 1997.

\bibitem{lennerz2018management}
Belinda~S Lennerz, Anna Barton, Richard~K Bernstein, R~David Dikeman, Carrie
  Diulus, Sarah Hallberg, Erinn~T Rhodes, Cara~B Ebbeling, Eric~C Westman,
  William~S Yancy, et~al.
\newblock Management of type 1 diabetes with a very low--carbohydrate diet.
\newblock {\em Pediatrics}, 141(6), 2018.

\bibitem{leohr2023ultra}
Jennifer Leohr, Mary~Anne Dellva, Elizabeth LaBell, David~E Coutant, Jorge
  Arrubla, Leona Plum-M{\"o}rschel, Eric Zijlstra, Tsuyoshi Fukuda, and Thomas
  Hardy.
\newblock Ultra rapid lispro (lyumjev{\textregistered}) shortens time to
  recovery from hyperglycaemia compared to humalog{\textregistered} in
  individuals with type 1 diabetes on continuous subcutaneous insulin infusion.
\newblock {\em Diabetes, Obesity and Metabolism}, 2023.

\bibitem{li2011hijacking}
Chunxiao Li, Anand Raghunathan, and Niraj~K Jha.
\newblock Hijacking an insulin pump: Security attacks and defenses for a
  diabetes therapy system.
\newblock In {\em 2011 IEEE 13th international conference on e-health
  networking, applications and services}, pages 150--156. IEEE, 2011.

\bibitem{loop}
Loop.
\newblock An automated insulin delivery app for ios, built on loopkit.
\newblock \url{https://github.com/LoopKit/Loop}.

\bibitem{loopkit}
LoopKit.
\newblock Tools for building automated insulin delivery systems on ios.
\newblock \url{https://github.com/LoopKit}.

\bibitem{marin2016feasibility}
Eduard Marin, Dave Singel{\'e}e, Bohan Yang, Ingrid Verbauwhede, and Bart
  Preneel.
\newblock On the feasibility of cryptography for a wireless insulin pump
  system.
\newblock In {\em Proceedings of the Sixth ACM Conference on Data and
  Application Security and Privacy}, pages 113--120, 2016.

\bibitem{moyer2012correlation}
James Moyer, Donald Wilson, Irina Finkelshtein, Bruce Wong, and Russell Potts.
\newblock Correlation between sweat glucose and blood glucose in subjects with
  diabetes.
\newblock {\em Diabetes technology \& therapeutics}, 14(5):398--402, 2012.

\bibitem{openaps}
OpenAPS.
\newblock The open artificial pancreas system project.
\newblock \url{https://github.com/openaps}.

\bibitem{panda2021secure}
Abhinandan Panda, Srinivas Pinisetty, and Partha Roop.
\newblock A secure insulin infusion system using verification monitors.
\newblock In {\em Proceedings of the 19th ACM-IEEE International Conference on
  Formal Methods and Models for System Design}, pages 56--65, 2021.

\bibitem{paul2011review}
Nathanael Paul, Tadayoshi Kohno, and David~C Klonoff.
\newblock A review of the security of insulin pump infusion systems.
\newblock {\em Journal of diabetes science and technology}, 5(6):1557--1562,
  2011.

\bibitem{ponder2019sugar}
Stephen~W Ponder and Kevin~L McMahon.
\newblock Sugar surfing in practice.
\newblock {\em AADE in Practice}, 7(6):26--30, 2019.

\bibitem{ponder2015sugar}
Stephen~W Ponder and Kevin~Lee McMahon.
\newblock {\em Sugar Surfing: How to manage Type 1 diabetes in a modern world}.
\newblock MediSelf Press Sausalito, CA, 2015.

\bibitem{reis2019site}
Charles Reis, Alexander Moshchuk, and Nasko Oskov.
\newblock Site isolation: Process separation for web sites within the browser.
\newblock In {\em 28th USENIX Security Symposium (USENIX Security 19)}, pages
  1661--1678, 2019.

\bibitem{roesner2013securing}
Franziska Roesner and Tadayoshi Kohno.
\newblock Securing embedded user interfaces: Android and beyond.
\newblock In {\em 22nd USENIX Security Symposium (USENIX Security 13)}, pages
  97--112, 2013.

\bibitem{rushanan2014sok}
Michael Rushanan, Aviel~D Rubin, Denis~Foo Kune, and Colleen~M Swanson.
\newblock Sok: Security and privacy in implantable medical devices and body
  area networks.
\newblock In {\em 2014 IEEE symposium on security and privacy}, pages 524--539.
  IEEE, 2014.

\bibitem{scheiner2020think}
Gary Scheiner.
\newblock {\em Think like a pancreas: A Practical guide to managing diabetes
  with insulin}.
\newblock Hachette Go, 2020.

\bibitem{seo2016flexdroid}
Jaebaek Seo, Daehyeok Kim, Donghyun Cho, Insik Shin, and Taesoo Kim.
\newblock Flexdroid: Enforcing in-app privilege separation in android.
\newblock In {\em NDSS}, 2016.

\bibitem{shi2019feedback}
Dawei Shi, Sunil Deshpande, Eyal Dassau, and Francis~J Doyle~III.
\newblock Feedback control algorithms for automated glucose management in t1dm:
  the state of the art.
\newblock {\em The artificial pancreas}, pages 1--27, 2019.

\bibitem{steiner2011optical}
Mark-Steven Steiner, Axel Duerkop, and Otto~S Wolfbeis.
\newblock Optical methods for sensing glucose.
\newblock {\em Chemical Society Reviews}, 40(9):4805--4839, 2011.

\bibitem{tang2010trust}
Shuo Tang, Haohui Mai, and Samuel~T King.
\newblock Trust and protection in the illinois browser operating system.
\newblock In {\em 9th USENIX Symposium on Operating Systems Design and
  Implementation (OSDI 10)}, 2010.

\bibitem{wang2009multi}
Helen~J Wang, Chris Grier, Alexander Moshchuk, Samuel~T King, Piali Choudhury,
  and Herman Venter.
\newblock The multi-principal os construction of the gazelle web browser.
\newblock In {\em USENIX security symposium}, volume~28, 2009.

\end{thebibliography}

\end{document}